\documentstyle[12pt]{article}

\begin{document}

\newcommand{\Jxtf}{x_{13} \cdot J(x_{12}) \cdot x_{24}}
\newcommand{\Jytf}{x_{13} \cdot J(w) \cdot x_{24}}
\newcommand{\Jxtt}{x_{13} \cdot J(x_{12}) \cdot x_{13}}
\newcommand{\Jxff}{x_{24} \cdot J(x_{12}) \cdot x_{24}}
\newcommand{\Jytt}{x_{13} \cdot J(w) \cdot x_{13}}
\newcommand{\Jyff}{x_{24} \cdot J(w) \cdot x_{24}}

\baselineskip=15.5pt

\begin{titlepage}

\begin{flushright}
hep-th/0002039\\
\end{flushright}
\vfil

\begin{center}
{\huge OPEs and 4-point Functions in AdS/CFT Correspondence
}
\end{center}

\vfil
\begin{center}
{\large Christopher P. Herzog}\\
\vspace{1mm}
Joseph Henry Laboratories, Princeton University,\\
Princeton, New Jersey 08544, USA\\
\vspace{3mm}
\end{center}

\vfil

\begin{center}
{\large Abstract}
\end{center}

\noindent
Recently 4-point correlation functions of axion and dilaton fields in type
IIB SUGRA on $\mbox{AdS}_5 \times \mbox{S}^5$ were computed
\cite{Freedman1}.  We reproduce from a CFT point of view all power law
singular terms in these AdS 4-point amplitudes.  We also calculate a
corresponding 4-point function in the weak coupling limit, $g_{YM}^2 N
\rightarrow 0$.  Comparison reveals the existence of a primary operator
that contributes to these same singular terms in the weak coupling limit
but which does not contribute to the power law singular terms of the type
IIB SUGRA 4-point functions.  We conclude that this new operator is not a
chiral primary and hence acquires a large anomalous dimension in the
strong coupling regime.

\vfil
\begin{flushleft}
January 2000
\end{flushleft}
\vfil
\end{titlepage}
\newpage


\section{Introduction}

The AdS/CFT correspondence relates string theories on anti-de Sitter space ($\mbox{AdS}_{d+1}$) backgrounds to $d$-dimensional conformal field theories (CFTs) \cite{Maldacena, Gubser, Witten, review}.  One simple example of this correspondence, which is the only example treated in this paper, is type IIB string theory on $\mbox{AdS}_5 \times \mbox{S}^5$ and ${\mathcal N} = 4, d = 4$ SU(N) Super Yang Mills (SYM) theory.  In the large N limit, with $g_{YM}^2 N$ held fixed and very large, the supergravity (SUGRA) approximation of type IIB string theory is valid, thus providing a perturbative way of understanding SYM theory at strong coupling.  

Correlation functions provide an important way of studying the correspondence.  Calculations of 2- and 3-point functions have already provided evidence that the correspondence is correct \cite{Freedman3, Lee, dHoker3}, but 4-point functions, as their form is not completely fixed by conformal invariance, can provide more detailed information about the CFT at strong coupling.  Previous studies of 4-point correlators which were peripherally useful in preparing this paper include \cite{Muck}-\cite{Intriligator}.

Recently in \cite{Freedman1}, for type IIB SUGRA on $\mbox{AdS}_5 \times \mbox{S}^5$, 
the first realistic 4-point functions
\[ 
\langle \phi(x_1) \phi(x_2) \phi(x_3) \phi(x_4) \rangle,\,  
\langle C(x_1) C(x_2) C(x_3) C(x_4) \rangle, 
\]
and
\[
\langle \phi(x_1) C(x_2) \phi(x_3) C(x_4) \rangle
\]
were calculated.\footnote{
These results were elaborated in \cite{dHoker1} where the original calculations were simplified and in \cite{dHoker2}
where the logarithmic singularities were explained.} The operators $\phi$ and $C$ correspond to the dilaton and axion supergravity fields.  As was pointed out in \cite{Gubser}, the dilaton and axion fields correspond to the operators $\phi \sim Tr(F^2 + \ldots)$ and $C \sim Tr(F \tilde{F} + \ldots)$ in ${\mathcal N}=4$ SYM theory.  We attempt to expand upon the results in \cite{Freedman1} by considering the corresponding CFT 4-point functions.

To make contact with CFT, it is convenient to expand these AdS 4-point functions as a power series
in $x_{13}^2$, $x_{24}^2$, and $x_{13}\cdot x_{24}$
in the ``direct'' or t-channel
limit $|x_{13}|, |x_{24}| \ll |x_{12}|$.\footnote{
Let $x_{ij} \equiv x_i - x_j$.}  
The power law singular terms in this series
are identical for all three 4-point functions and come solely from graviton
exchange.

Because of the proposed AdS/CFT correspondence, we expect
to be able to reconstruct the AdS 4-point amplitudes in terms of a double operator product expansion (OPE) in CFT.  We express
$\phi(x)\phi(y)$ and $C(x) C(y)$ in terms of their OPEs.  Multiplying two such
OPEs together and taking the 2-point functions of operators in the expansion
should recover the scattering amplitude.  
In the limit where $|x_{13}|$ and $|x_{24}|$ are very small compared to
$|x_{12}|$, we expect to be able to represent the 4-point amplitude schematically as
\begin{equation}
\langle {\mathcal{O}}_1(x_1) {\mathcal{O}}_2(x_2) 
{\mathcal{O}}_3(x_3) {\mathcal{O}}_4(x_4) \rangle =
\sum_{n,m} \frac{\alpha_n \langle 
{\mathcal{O}}_n (X_{13}) {\mathcal{O}}_m (X_{24}) \rangle \beta_m}
{x_{13}^{\Delta_1+\Delta_3-\Delta_m} x_{24}^{\Delta_2+\Delta_4-\Delta_n}}
\label{OPE}
\end{equation}
where ${\mathcal{O}}_p$ is some operator of
dimension $\Delta_p$.  We have defined
$X_{ij} \equiv (x_i + x_j)/2$.  The expression
is schematic because in general the
operator ${\mathcal{O}}_p$ may be a tensor.

Because the leading order terms in the 
three 4-point functions come from graviton
exchange and because of the proposed 
AdS/CFT correspondence, we expect and it was
shown in \cite{Freedman1} that the leading order term in (\ref{OPE}) comes from
the 2-point function of the energy momentum tensor with itself, $\langle
T_{ab}(X_{13}) T_{cd}(X_{24}) \rangle$.  In this paper, we go further and show
precisely how, from a CFT point of view, 
all singular terms in the t-channel limit
of the three 4-point functions arise from 
exchange of $T_{ab}$ and its descendants.  We also investigate how the 4-point functions change as we move from strong to weak coupling.

The work proceeds in four parts.  First we look at the leading order singular
terms in the AdS 4-point functions in 
the t-channel limit, the same terms that in
the next section we will be able to compute using conformal invariance.  In the
third and fourth sections, we compute the equivalent 4-point function in the weak coupling limit of the CFT, which is essentially the case of electricity and magnetism, in order to try to understand how the 4-point function changes as we move from strong to weak coupling.  This
investigation will reveal the existence of a new nonchiral primary operator in the weak limit which, if the AdS/CFT correspondence is to hold, acquires a large anomalous dimension as we move to strong coupling and hence does not contribute to the AdS singular terms.

\section{AdS 4-point Functions}

It turns out that in the t-channel limit, all three of the AdS 4-point functions calculated in \cite{Freedman1} have the same singular power law terms.  Moreover, the singular terms come only from t-channel graviton exchange.  The singular terms in the amplitude are
\begin{equation}
\left. I_{\mbox{\scriptsize grav}} \right|_{\mbox{\scriptsize sing}} = 
\frac{2^{10}}{35 \pi^6} 
\frac{1}{x_{13}^8 x_{24}^8} \left[ s(7t^2 + 6t^4) + s^2(-7+3t^2)-8s^3 \right].
\label{singular}
\end{equation}
The variables $s$ and $t$ are conformally invariant functions of the $x$'s:
\begin{eqnarray}
\label{s}
s & \equiv & \frac{1}{2} \frac{x_{13}^2 x_{24}^2}
{x_{12}^2 x_{34}^2 + x_{14}^2 x_{23}^2} \, ,\\
\label{t}
t & \equiv & \frac{x_{12}^2 x_{34}^2 - x_{14}^2 x_{23}^2}
{x_{12}^2 x_{34}^2 + x_{14}^2 x_{23}^2} \, .
\end{eqnarray}

To make contact with (\ref{OPE}), we can expand $I_{\mbox{\scriptsize grav}}$ in one of two ways.  We can carry out an asymmetric expansion in powers of $x_{12}$ or we can perform a symmetric expansion in powers of $w \equiv (X_{13} - X_{24})$.  We consider only the symmetric expansion because terms odd in powers of $w$ will not appear:

\begin{eqnarray*}
s & = & \frac{x_{13}^2 x_{24}^2}{4 w^4} 
\frac{1}{g(w, x_{13}, x_{24})} \, \, ,\\
t & = & -\frac{\Jytf}{w^2}
\left[ 1 + \frac{1}{4} \frac{x_{13} \cdot x_{24}}{w^2}
\frac{x_{13}^2 + x_{24}^2}{\Jytf}
\right]
\frac{1}{g(w, x_{13}, x_{24})}\, \, .
\end{eqnarray*}
In the above, $J_{ij} = \delta_{ij} - 2x_i x_j / x^2$ is the Jacobian tensor of the conformal inversion $x_i' = x_i/x^2$, and we have defined
\begin{eqnarray*}
g(w, x_{13}, x_{24}) & \equiv & 
\left[ 1 + \frac{1}{2w^2}(x_{13}^2 + x_{24}^2) 
- \frac{1}{w^4}((w \cdot x_{13})^2 + (w \cdot x_{24})^2) + \right.\\
& & \left.+ \frac{1}{16 w^4}(x_{13}^4 + x_{24}^4 + 
2x_{13}^2 x_{24}^2 + 4(x_{13}\cdot x_{24})^2) 
\right].
\end{eqnarray*}

Armed with these expressions for $s$ and $t$, we can expand $I_{\mbox{\scriptsize grav}}$ to the third nontrivial order in $w$, i.e. we consider terms of order $w^{-n}$ where 
$n = 8, \, 10, \, \mbox{and} \, 12$. 
The amplitude can be written order by order as
\begin{equation}
\left. I_{\mbox{\scriptsize grav}} 
\right|_{8^{\mbox{\scriptsize th}}} 
 =  \frac{2^6}{5 \pi^6} \frac{1}{x_{13}^6 x_{24}^6 w^8}
\left[ 4 (\Jytf)^2 - x_{13}^2 x_{24}^2 \right] ,
\label{grav8}
\end{equation}
\begin{eqnarray}
\lefteqn{
\left. I_{\mbox{\scriptsize grav}} 
\right|_{10^{\mbox{\scriptsize th}}}
 = 
\frac{2^6}{5 \pi^6} \frac{1}{x_{13}^6 x_{24}^6 w^{10}}
\left[ \right. } \nonumber \\
& &-6 (\Jytf)^2 \left[
(\Jytt) + (\Jyff) \right] + \nonumber \\
& & + 2 x_{13} \cdot x_{24} (x_{13}^2 + x_{24}^2) 
(\Jytf) + \nonumber \\
& & \left. + x_{13}^2 x_{24}^2 \left[ (\Jytt) + (\Jyff) \right] \right], 
\label{grav10}
\end{eqnarray}
\begin{eqnarray}
\lefteqn{
\left. I_{\mbox{\scriptsize grav}} 
\right|_{12^{\mbox{\scriptsize th}} \mbox{\scriptsize , asym}} 
 = 
\frac{2^6}{5 \pi^6} \frac{1}{x_{13}^6 x_{24}^6 w^{12}}
\left[ \right.} \nonumber \\
& & 6 (\Jytt)^2 (\Jytf)^2+
\nonumber \\
& & -3 x_{13}^2 (x_{13}\cdot x_{24}) (\Jytt) (\Jytf) +\nonumber \\
& & - \frac{3}{4} x_{13}^4 (\Jytf)^2 -\frac{3}{4} x_{13}^2 x_{24}^2 (\Jytt)^2 +
\nonumber \\
& &\left.
+ \frac{1}{4} (x_{13}\cdot x_{24})^2 x_{13}^4+\frac{1}{8} x_{13}^6 x_{24}^2 + (x_{13} \leftrightarrow x_{24})\right],
\label{gravasym}
\end{eqnarray}
\begin{eqnarray}
\lefteqn{
\left. I_{\mbox{\scriptsize grav}} 
\right|_{12^{\mbox{\scriptsize th}} \mbox{\scriptsize , sym}} 
 = 
\frac{2^6}{5 \pi^6} \frac{1}{x_{13}^6 x_{24}^6 w^{12}}
\left[ \right. } \nonumber \\
& & + \frac{24}{7} (\Jytf)^4+12 (\Jytf)^2 (\Jytt)(\Jyff) +\nonumber \\
& & -\frac{3}{2} x_{13}^2 x_{24}^2 (\Jytt)(\Jyff) + \nonumber \\
& & - 3 x_{13}\cdot x_{24} (\Jytf)
\left[ x_{13}^2 (\Jyff) + x_{24}^2 (\Jytt)\right] + \nonumber \\
& &-3 (\Jytf)^2 (\frac{5}{14} x_{13}^2 x_{24}^2 + (x_{13}\cdot x_{24})^2) +
\nonumber \\
& & \left.+ x_{13}^2 x_{24}^2 (x_{13}\cdot x_{24})^2 
- \frac{1}{28} x_{13}^4 x_{24}^4 
 \right].
\label{gravsym}
\end{eqnarray}
The abbreviations sym and asym in the previous two equations symbolize that we
have split the twelfth order term into pieces with equal and unequal numbers of
$x_{13}$ and $x_{24}$ respectively.  As these two pieces come from different
2-point functions in the double OPE (\ref{OPE}), this separation will be
important. Note that the $st^4$, $s^2t^2$, and $s^3$ terms contribute only to 
the symmetric twelfth order term, (\ref{gravsym}).

\section{A General CFT}

As is well known, conformal invariance alone does not completely specify the form
of a 4-point function.  For two pairs of scalars of dimension four, the most we
can say is that
\[
\langle {\mathcal O}'(x_1) {\mathcal O}(x_2) 
{\mathcal O}'(x_3) {\mathcal O}(x_4)\rangle  =
\frac{1}{x_{13}^8 x_{24}^8} F(s, t)
\]
where $F$ is some unknown function of the conformally 
invariant variables $s$ and
$t$.  If we know the primary operators that occur in the OPEs of $\mathcal O O$
and $\mathcal O' O'$ along with their coefficients, then we can specify $F$
completely.  As was shown in \cite{Freedman1}, the assumption that $T_{ab}$
appears in the OPE reproduces the leading order term in the 
4-point function in the
t-channel limit, (\ref{grav8}).  We will show that this assumption actually
reproduces all the terms in (\ref{singular}).

In the CFT literature, equations have been derived that describe the conformal block contribution, up to an overall normalization factor, of an arbitrary
tensor primary operator exchanged in a 4-point interaction of four arbitrary
scalars.  We have tried unsuccessfully to use Eq. 3.15 of \cite{Ferrara} and suspect there may be some normalization problem.  A similar equation can be found in \cite{Ruhl} but appears to be more difficult to apply and was discovered only after the following work had been completed. 

We shall use a brute force 
approach that has the advantage of showing us
precisely how $T_{ab}$ and its descendants arise in the OPE of $\mathcal O O$.  Our approach is similar to methods for calculating 4-point functions that can be found in the CFT literature \cite{Petkou}.  
We may write the symmetric OPE schematically as 
\begin{eqnarray}
\lefteqn{
{\mathcal O}\left(\frac{x}{2}\right) {\mathcal O} \left(-\frac{x}{2}\right) \sim
} \nonumber \\
& & A \frac{x_a x_b}{x^6} \left[ T_{ab} (0) - 
\frac{1}{2} x_i x_j T^{(2)}_{abij}(0) +
\frac{1}{24} x_i x_j x_k x_l T^{(4)}_{abijkl}(0)
+ \ldots \right]
\label{scalarOPE}
\end{eqnarray}
where $A$ is an overall constant and
$T^{(2)}_{abij}$ and $T^{(4)}_{abijkl}$ are second and fourth order
descendants of $T_{ab}$.  From conformal 
invariance, we know that the descendants
can be expressed as derivatives of $T_{ab}$.  
More specifically, we can write the
second order descendant as
\begin{equation}
T^{(2)}_{abij}(x)  =  \mu \partial_i \partial_j T_{ab}(x) 
+ \nu \delta_{ij} \Box T_{ab}(x),
\label{Ttwo}
\end{equation}
and the fourth order descendant can be written 
correspondingly in terms of fourth
order derivatives of $T_{ab}$.
Terms of odd order in $x$ would be inconsistent with
the symmetry of the expansion.  

As a warm up, we consider the contribution of $T_{ab}$ alone to the 4-point
function.  The leading nontrivial
term in the OPE of two scalars is by assumption $T_{ab}$ and as $T_{ab}$ has
dimension 4, the two point function of $T_{ab}$ with itself that appears inside
the 4-point function must have a $w^8$ in the denominator.  In other words,
the leading term in the power series $F$ must be of the form $\alpha s^2 + \beta
st^2 + \gamma t^4$.

In fact we can do better.  From conformal invariance (see for example
\cite{Osborn}), we know that the 2-point function of $T_{ab}$ with itself must be
\begin{equation}
\langle T_{ab}(x_1) T_{cd}(x_2) \rangle =
\frac{C_T}{x_{12}^8} [J_{ac}(x_{12}) J_{bd}(x_{12})
+ J_{ad}(x_{12}) J_{bc}(x_{12}) - \frac{2}{d} \delta_{ab} \delta_{cd}].
\label{emttwo}
\end{equation}
We can at this stage show agreement between the CFT and the AdS 4-point functions
at leading order, as was done in \cite{Freedman1}.  Using (\ref{scalarOPE}) and
(\ref{emttwo}), we see that (\ref{OPE}) agrees with the leading order term
(\ref{grav8}) provided $A^2 C_T = 2^7 / (5 \pi^6)$.  Another way of
understanding this calculation is to say that (\ref{emttwo}) is consistent with
$F$ only if $\alpha = -\beta$ and if $\gamma = 0$. Note that in $F$, terms odd in
powers of $t$ are not allowed as they are not consistent with the structure of the
symmetric OPE of two scalars.  Therefore, next order terms will be of the form
$s^3$, $s^2 t^2$, $s t^4$, and possibly $t^6$.
 
To proceed further in our calculation of the double OPE, note that one class of 2-point functions that need to be calculated involve $T_{ab}$ 
in one OPE with $T_{ab}$ and any of its descendants
in the other OPE.  We can obtain this entire class easily from \cite{Osborn} or \cite{Ferrara}, where the authors use conformal invariance to show
that the 3-point function of two scalars with $T_{ab}$ must take the form
\[
\langle T_{ab}(x_1) {\mathcal O}(x_2) {\mathcal O}(x_3) \rangle =
\frac{a}{x_{12}^d x_{13}^d x_{23}^{2\eta - d}} t_{ab}(X_{23})
\]
where
\[
t_{ab}(X) = \left( \frac{X_a X_b}{X^2} - \frac{1}{d} \delta_{ab} \right)
\]
and where
\begin{eqnarray*}
X_{23} &=& \frac{x_{21}}{x_{21}^2} - \frac{x_{31}}{x_{31}^2} \, ,\\
X_{23}^2 & = & \frac{x_{23}^2}{x_{21}^2 x_{31}^2}.
\end{eqnarray*}
In the above expression, $a$ is an as yet undetermined constant, $d$ is the
dimension of space, and $\eta$ is the dimension of ${\mathcal O}$. 
To read off the
2-point functions of interest, we consider the limit $x_{23} \approx 0$, and we
expand the three point function in the variables $x = x_{23}$ and $y = (x_{12} +
x_{13})/2$.  In the case where $d = 4 = \eta$, the resulting somewhat cumbersome
expression for the 3-point function is
\begin{eqnarray}
\lefteqn{\langle T_{ab}(x_1) {\mathcal O}(x_2) {\mathcal O}(x_3) \rangle =
\frac{a}{x^6 y^{12}} \left( 1 + \frac{x^2}{2y^2} - \frac{(x \cdot y)^2}{y^4}
+ \frac{x^4}{16 y^4} \right)^{-3} } \nonumber \\
& & \left[ 4 (x \cdot y)^2 y_a y_b - 
x \cdot y (2y^2 +\frac{1}{2} x^2) 
(x_a y_b + x_b y_a) + \right. \nonumber \\
& & + (y^4 + \frac{1}{2} x^2 y^2 + \frac{1}{16}x^4) x_a x_b +\nonumber \\
& & \left. -\frac{1}{4} \delta_{ab} x^2 (y^4 + \frac{1}{2} x^2 y^2
-(x \cdot y)^2 + \frac{1}{16} x^4) \right]
\label{threept}
\end{eqnarray}

Now the lowest order term in the above expression corresponds to the 2-point
function of $T_{ab}$ with itself, which was discussed previously. As a check on
the calculations so far, one may verify that (\ref{emttwo}) is completely
consistent with the highest order term in (\ref{threept}) if $A C_T = a/2$.

The second order term in (\ref{threept}) corresponds to the 2-point function of
$T_{ab}$ with the descendant operator $T^{(2)}_{abij}$.  One finds that up to
permutations of the indices $(abij)$, the two point function takes the form
\begin{eqnarray}
\langle T^{(2)}_{abij}(x) T_{cd}(0) \rangle & = & 
\frac{-2C_T}{x^{10}} \left[ -3 J_{ac} J_{bd} J_{ij} + \right. \nonumber \\
& & \left. + \frac{1}{2} \delta_{ij}(\delta_{ac} J_{bd} + \delta_{bd} J_{ac}) + 
\frac{1}{2} \delta_{ab} \delta_{cd} J_{ij} \right]
\label{desctwo}
\end{eqnarray}
where all the $J$ take $x$ as an argument. 
Given the same condition on $A$ and $C_T$
as above, we have agreement between (\ref{threept}) and the higher order term
(\ref{grav10}) in the 4-point function.  

The third order term in (\ref{threept}) corresponds to the 2-point function of
$T_{ab}$ with the descendant operator $T^{(4)}_{abijkl}$.  
We have indeed checked
that this third order term agrees with the 
asymmetric term (\ref{gravasym}) in the
4-point function given the same conditions on $A$ and $C_T$. 

Note that (\ref{desctwo}) is consistent with (\ref{Ttwo}) only if $\mu = -1/28$
and $\nu = 1/28$.  Calculating the 2-point function of $T^{(2)}_{abij}$ with
itself then becomes a simple matter of taking derivatives of (\ref{emttwo}).  We
have checked that 
$\langle T^{(2)}_{abij} T^{(2)}_{cdkl} \rangle$ agrees with the
symmetric term (\ref{gravsym}) in the 4-point function.  This calculation fixes
the coefficients of $st^4$, $s^2 t^2$, and $s^3$ and also
shows that $t^6$ does not appear in the singular terms.  

There are other types of expansions one could use to compare the AdS results with CFT.  For example, one could have taken a limit in which only two scalars approach one another.  Then, instead of 2-point functions, one considers the set of 3-point functions involving the two other scalars and the operators in the OPE of the two neighboring scalars.  We have followed the lead of \cite{Freedman1} and used the double OPE method.

\section{${\mathcal N} = 4$ SYM at Weak Coupling}

Conformal invariance implies that the coordinate dependence of the 2- and 3-point functions in AdS/CFT correspondence does not change as we move from weak to strong coupling.  In addition, nonrenormalization theorems are thought to keep the coefficients of 2- and 3-point correlation functions involving chiral primaries independent of coupling \cite{review}.  However, the coordinate dependence of 4-point functions can and does change as we vary the coupling.  Thus a calculation and comparison of 4-point functions in the two coupling regimes is likely to be a much more enlightening way of seeing how the theory changes as we move from weak to strong coupling.  So far, we have only looked at the strong coupling regime.

In this section, we begin a consideration of ${\mathcal N} = 4$ SYM theory in the weak coupling limit which will culminate in the next section with a computation of the connected dilaton 4-point function at weak coupling to leading order in $\lambda = g_{YM}^2 N$.  Essentially, this correlation function is equivalent to the 4-point function of $F^2$ in electricity and magnetism as the difference between the two only appears at subleading order in $\lambda$.  

To be more specific, the dilaton and axion operators take the form $\phi \sim Tr(F^2 + \ldots)$ and $C \sim Tr(F \tilde{F} + \ldots)$.\footnote{
To be completely precise, 
\[\tilde{F}_{ab} = \frac{1}{2} \epsilon_{abcd} F_{cd} \, . 
\]}
As shown in \cite{Klebanov} in the case of the dilaton, the higher order terms will involve three or more of the operators $F^{kl}_{ab}, X^{kl}_a$, and $\Theta^{kl}_\alpha$.\footnote{$k, l, \ldots$ are SU(N) indices, $a, b, \ldots$ are spatial indices, and $\alpha, \beta, \ldots$ are spinor indices.}  

The 2-point function for
$F_{ab}$ is:
\begin{equation}
\langle F^{kl}_{ab}(x_1) F^{mn}_{cd}(x_2) \rangle = 
\frac{c}{x_{12}^4} \delta^{kn} \delta^{lm} [J_{ac}(x_{12}) J_{bd}(x_{12}) 
- J_{ad}(x_{12}) J_{bc}(x_{12}) ]
\label{Ftwo}
\end{equation}
where $J_{ab}$ is as defined above and $c \sim g_{YM}^2$ is a constant.  In general, the two point function of an operator with itself will contain these same Kronecker delta functions of the SU(N) indices.  From this fact and Wick's Theorem, it is not difficult to see that the higher order terms in $\phi$ and $C$ involving three or more operators produce corrections to the correlation functions which are higher order in $g_{YM}^2 N$.  From hereon, we suppress the SU(N) indices and consider only the leading order terms in $\phi$ and $C$.

As described in the introduction, scattering in the t-channel limit can be represented in terms of a double OPE.  Thus, first we express
$\phi(x) \phi(y)$ and $C(x) C(y)$ in terms of their OPEs. 
It turns out that up to terms with no
contractions, the dilaton and axion have the same OPE.  We present first an
intermediate result:
\begin{eqnarray}
\lefteqn{F^2(x) F^2(0) \sim F \tilde{F} (x) F \tilde{F} (0) \sim} 
\nonumber \\
& & \frac{48 c^2}{x^8} - \frac{32 c x_a x_b}{x^6} \left[ F_{ac}(x) F_{bc}(0) - 
\frac{1}{4} \delta_{ab} F_{cd}(x) F_{cd}(0) \right].
\label{twodil}
\end{eqnarray}
This expression is reminiscent of the energy momentum tensor which takes the form 
\[
T_{ab}(x) = K [ F_{ac} F_{bc} - \frac{1}{4} \delta_{ab} F^2](x)
\label{emt}
\]
where $K$ another constant.\footnote{
The constants in this section are related to those in the 
previous section by $C_T = 2 c^2 K^2$ and $A = 32 c / K$.}
Thus, the symmetric OPE of the dilaton and axion can be written as
\begin{eqnarray}
\lefteqn{F^2 \left 
 (\frac{x}{2} \right) F^2 \left(-\frac{x}{2} \right) \sim
\frac{48 c^2}{x^8} \, +}  
\nonumber \\ 
 & & -\frac{32 c x_a x_b}{K x^6} \left[
T_{ab}(0) - \frac{1}{2} x_i x_j (T^{(2)}_{abij}(0) + P_{abij}(0))
\right]
+ \ldots
\label{diltwos}
\end{eqnarray}
where the second order descendant $T^{(2)}_{abij}$ takes the same form as in
(\ref{Ttwo}) and we have found potentially a new primary with the complicated
form
\begin{equation}
P_{abij}(x) \equiv 
-\frac{3}{14} \partial_i \partial_j T_{ab} (x) -
\frac{1}{28} \delta_{ij} \Box T_{ab}(x) +
T'_{abij}(x)
\label{newop}
\end{equation}
where
\[
T'_{abij}(x) \equiv
 K \left[ (\partial_i F_{ac}) (\partial_j F_{bc})(x)
- \frac{1}{4} \delta_{ab} (\partial_i F_{cd})(\partial_j F_{cd})(x) \right].
\]
Using Wick's Theorem, we have checked that the 2-point function of $P_{abij}$
with $T_{ab}$ and with $F^2$ is zero.  In
the next section, we will see from evaluating the dilaton 4-point function,
that the 2-point function of $P_{abij}$
with itself is nonzero, and therefore that
$P_{abij}$ does not vanish by the equations of motion.  To show definitively
that $P_{abij}$ is a primary operator, it would be nice to demonstrate that
it transforms appropriately under the conformal group and more specifically
under inversion.  Preliminary results suggest that $P_{abij}$ does transform
as a primary under inversion, but the full calculation is lengthy and has
not been completed.  From the index structure, it is clear that $P_{abij}$
has spin four, and at least in the weak coupling regime, the dimension of
this new primary operator is six.

If the operator $P_{abij}$ were a chiral primary, its dimension would be
algebraically protected, and the operator should contribute to the 4-point
functions equally at weak and strong coupling.  However, as we have seen,
$P_{abij}$ does not contribute to the leading singular terms at strong
coupling and, as we will see in the next section, $P_{abij}$ does contribute
at weak coupling.  Moreover, our new primary does not correspond to any of
the known chiral primaries of dimension six.  The logical conclusion is that
$P_{abij}$ is nonchiral and acquires a large anomalous dimension in the
strong coupling regime:  That there seem to be no nonchiral fields on AdS
space with a mass below the string scale suggests that our nonchiral primary
has a dimension which grows at least as fast as $(g_{YM}^2 N)^{1/4}$ in the
strong coupling limit, $g_{YM}^2 N \rightarrow \infty$ \cite{review}.

\section{The Dilaton 4-point Function in the Weak Coupling Limit of SYM}

We calculate the four dilaton amplitude 
in the weak coupling limit of ${\mathcal N} = 4, d= 4$ SYM theory, which as noted in the previous section, is essentially equivalent, at leading order in $g_{YM}^2 N$, to the 4-point function of $F^2$ of electricity and magnetism, i.e.
\[
M_4 \equiv \langle F^2(x_1) F^2(x_2) F^2(x_3) F^2(x_4) \rangle.
\]
Let 
\[
W_{1324} \equiv \langle F_{ab}(x_1) F_{cd}(x_3) \rangle
\langle F_{cd}(x_3) F_{ef}(x_2) \rangle
\langle F_{ef}(x_2) F_{gh}(x_4) \rangle
\langle F_{gh}(x_4) F_{ab}(x_1) \rangle
\]
and let
\[
W_{12} \equiv \langle F_{ab}(x_1) F_{cd}(x_2) \rangle
\langle F_{cd}(x_2) F_{ab}(x_1) \rangle =
\frac{24 c^2}{x_{12}^8}.
\]
Then
\[
M_4 = 16 (W_{1234} + W_{1324} + W_{1342}) 
+ 4(W_{12} W_{34} + W_{13} W_{24} + W_{14} W_{23}).
\]
The terms containing $W_{ab}$ do not concern us as they describe the disconnected
pieces of the 4-point function.  Note that by definition $W_{abcd} = W_{adcb}$.  We proceed with a calculation of the amplitude
$W_{1324}$.  If we define
\[
A_{ab} \equiv J_{ac}(x_{13})J_{ce}(x_{23})J_{eg}(x_{24})J_{gb}(x_{14}),
\]
then the amplitude takes the form
\[
W_{1324} = \frac{8 c^4}{x_{13}^4 x_{24}^4 x_{23}^4 x_{14}^4}
[(\mbox{tr} A)^2 - \mbox{tr} A^2].
\]
Let $\lambda_i$ be the four eigenvalues of $A$.  Clearly
\[
Ch(A) \equiv (\mbox{tr} A)^2 - \mbox{tr} A^2 = 2 \sum_{i<j} \lambda_i \lambda_j.
\]
A brief consideration of $Ch(A)$ reveals that it is invariant under the conformal
group.  In particular, we consider the case in which we use a translation to set
$x_3 = 0$ and then an inversion to send it off to infinity.  In this case,
\[
J_{ac}(x_{13}) J_{ce}(x_{23}) = \delta_{ae} + O(|x_3|^{-1}).
\]
If we then choose a basis in which $x_{24} = (a, 0, 0, 0)$ and $x_{14} = (b \sin
\phi, b \cos \phi, 0, 0)$, the matrix $A$ becomes effectively two dimensional:  
Two of the $\lambda_i$ equal one, and the remaining two can be obtained by
diagonalizing the matrix
\[
\left( \begin{array}{cc}
-\cos 2\phi & -\sin 2 \phi \\
\sin 2 \phi & -\cos 2 \phi
\end{array} \right)
\]
giving $\lambda_\pm = - \exp(\pm 2 \phi i)$.  Hence
\[
Ch(A) = 4 ( -1 + 4 \sin \phi)\, .
\]

As $Ch(A)$ is invariant under the conformal group, it must be expressible as a
function of $s$ and $t$.  In the limit $x_3 \rightarrow \infty$
\begin{eqnarray*}
s & \rightarrow & \frac{1}{2} \frac{x_{24}^2}{x_{12}^2 + x_{14}^2} ,\\
t & \rightarrow & \frac{ x_{12}^2 - x_{14}^2}{ x_{12}^2 + x_{14}^2}.
\end{eqnarray*}
It is now a straightforward matter to express $Ch(A)$ in terms of $s$ and $t$:
\[
Ch(A) = \frac{4}{s(1-t)} (-s + t^2 -3st +4s^2).
\]
Indeed, Mathematica was used to verify that this expression is correct.
It follows immediately that
\[
W_{1324} = \frac{2^9 c^4}{x_{13}^8 x_{24}^8} \frac{s(-s+t^2-3st+4s^2)}{(1-t)^3}.
\]
Repeating the calculation for the other two $W_{abcd}$, we obtain the connected
4-point function
\begin{eqnarray*}
\left. M_4 \right|_{\mbox{\tiny connected}} &=& 
\frac{2^{14} c^4}{x_{13}^8 x_{24}^8}
\frac{1}{(1-t^2)^3} [ s(-s + t^2 + 4s^2 - 12st^2 + 3t^4 + 12s^2 t^2 - 3st^4) +\\
& & + 2^3 s^4 (3 - 16s + t^2 + 16s^2) ].
\end{eqnarray*}
As one can see, the leading order terms $s^2$ and $st^2$ are in agreement with
(\ref{singular}). However, the
coefficients of the higher
order terms are quite different from those in (\ref{singular}), thus
demonstrating that a new primary, or primaries, 
appear at this level, as we indeed saw in the previous section.

\section{Discussion, Conclusions, and Ideas for Future Work}

We have successfully reproduced all of the singular terms in the four
dilaton, four axion, and two dilaton-two axion 4-point functions
calculated in \cite{Freedman1} using only the assumptions of conformal
invariance and the presence of $T_{ab}$ in the OPE of two scalars.  
Moreover, we have developed a better understanding of the structure of the
descendants of $T_{ab}$ in this OPE.  Comparison of this strongly coupled
result to the weakly coupled limit revealed the presence of a nonchiral
primary $P_{abcd}$ (see equation \ref{newop}).  At weak coupling, this
nonchiral primary has spin four and dimension six.  However, this primary
is believed to have an anomalous dimension which grows at least as fast as
$(g_{YM}^2 N)^{1/4}$, thus effectively disappearing as an exchanged
operator in the 4-point functions calculated by \cite{Freedman1}.

Some lines of inquiry remain open.  It would be nice to know the exact form
of $T^{(4)}_{abcdef}$, and knowing this form would allow the work done here
to be extended to the order $n=16$, the first order in the strongly coupled
4-point functions calculated by \cite{Freedman1} where new chiral primaries
are expected to appear.  The difference between the conformal block
contribution of $T_{ab}$ and the correlation functions as calculated by
\cite{Freedman1} would then presumably give us some insight as to the
precise nature of these additional operators, allowing, perhaps, a better
understanding of the logarithms that appear in these 4-point functions.

The present method of calculating descendants order by order becomes
extremely cumbersome to apply at higher order, so a more efficient approach
may be to use variants of equations given in \cite{Ferrara} and \cite{Ruhl}.  
Although, as was mentioned above, we have had trouble using Eq.~3.15 of
\cite{Ferrara} directly, we do have a guess as to how to modify the
equation, and the modified version matches the AdS results in a way we would
expect.

\section*{Note Added for Publication}

A year has elapsed between the writing of this paper and its submission
for publication.  During the interim several papers have appeared which
use results derived here and which answer some of the questions raised in
this paper.  For example, in \cite{Dolan}, more explicit formulae for the
conformal block contribution of higher spin operators to scalar four point
functions were derived.  These results hopefully clarify the confusion in
this paper concerning the work of \cite{Ferrara}.  Another important work
is \cite{Hoffman} where, using some results derived here, the program
suggested in the conclusion of this paper was successfully carried out.

\section*{Appendix}

Here are some useful identities involving the tensor
\[
J_{ab}(x) \equiv \delta_{ab} - \frac{2 x_a x_b}{x^2}.
\]
In what follows, the argument of $J_{ab}$ will be $x$.  
First, here are some elementary properties of the tensor:
\[
J_{ab} J_{bc} = \delta_{ac} \;\; ; \;\; 
J_{aa} = 2 \;\; ; \;\;
x_a J_{ab} = - x_b \, .
\]
Next, here are some trivial index rearrangements:
\[
x_i J_{ab} = x_b J_{ia} + x_i \delta_{ab} - x_b \delta_{ia} \, ,
\]
\[
J_{cd} J_{ij} = J_{ic} J_{jd} + \delta_{ij} J_{cd} + \delta_{cd} J_{ij}
- \delta_{ic} J_{jd} - \delta_{jd} J_{ic} + \delta_{ic} \delta_{jd}
- \delta_{cd} \delta_{ij} \, .
\]
Finally, here are some derivatives of $J_{ij}$:
\[
\partial_m J_{ij} = - \frac{2}{x^2} ( 
x_m J_{ij} - x_m \delta_{ij}
+ x_i \delta_{jm} + x_j \delta_{im} ),
\]
\begin{eqnarray*}
\partial_n \partial_m J_{ij} &=& - \frac{2}{x^2} ( 2 J_{mn} J_{ij}
-2 \delta_{ij} J_{mn} - \delta_{mn} J_{ij} +\\
& &  + \delta_{jm} J_{in} +
\delta_{im} J_{jn} + \delta_{jn} J_{im} + \delta_{in} J_{jm} + \\
& & -\delta_{im} \delta_{jn} - \delta_{jm} \delta_{in} + 
\delta_{mn} \delta_{ij}).
\end{eqnarray*}

\section*{Acknowledgments}

The author would like to acknowledge many useful discussions with Igor
Klebanov.  He is also grateful for several useful conversations with
Leonardo Rastelli.  Thanks also go to A. Petkou, who, after the first
electronic submission of this paper, brought to our attention relevant
literature concerning O(N) non-linear $\sigma$-models.

\end{document}